\documentclass[aps,pra,twocolumn,superscriptaddress]{revtex4}
\usepackage{enumerate}
\usepackage{graphicx}
\usepackage{multirow,amssymb,amsbsy,amsmath}
\usepackage{bm}
\newcommand*{\ket}[1]{\mathopen{|}#1\mathclose{\rangle}}
\newcommand*{\bra}[1]{\mathopen{\langle}#1\mathclose{|}}
\begin{document}


\title{Extraction of Pure Entangled States from Many Body Systems by Distant Local Projections}


\author{ J. Molina }
\affiliation{Department of Systems Engineering and Automation. Technical University of Cartagena. 30202. Cartagena. Spain}
\author{ H. Wichterich  }
\affiliation{ Department of Physics and Astronomy. University College London. Gower Street, London WC1E 6BT, UK }

\author{ V.E. Korepin}
\affiliation{C.N. Yang Institute for Theoretical Physics, State University of New York at Stony Brook. Stony Brook NY 11794-3800 USA}

\author{ S. Bose  }
\affiliation{ Department of Physics and Astronomy. University College London. Gower Street, London WC1E 6BT, UK }


\date{\today}

\begin{abstract}
We study the feasibility of extracting a pure entangled state of
non-complementary, and potentially well separated, regions of a
quantum many-body system. It is shown that this can indeed be
accomplished in non-equilibrium scenarios as well as the ground state
of the considered spin chain models when one {\em locally} measures
observables such as magnetization in separated blocks of spins. A
general procedure is presented, which can search for the optimal way
to extract a pure entangled state through local projections. Our
results indicate a connection of the projective extraction of
entanglement to good quantum numbers of the underlying Hamiltonian.
\end{abstract}


\maketitle

\section{Introduction}\label{sec.intro}
Much attention as been paid to entanglement, the purely {\em quantum} part of correlations, since the emergence of the quantum information science. In this context, quantum entanglement can be regarded as a resource for quantum information processors \cite{Bennett2000}. As example of this assertion, entanglement is believed to be the major resource for the acceleration capabilities of quantum processors in performing computation and communication tasks such as teleportation or superdense coding. With this in mind, it is important to note that a typical quantum processor is also a quantum many body system. Thus, the attention that quantum information community has paid in the last years to quantum many body systems intensively studied in condensed matter physics has lead to consider that condensed matter phenomena can be explored from an information theoretic point of view, particularly studying entanglement in quantum many body systems \cite{Osterloh2008} and using these systems for fruitful operations in quantum communications \cite{Bose2007} and computation protocols \cite{Yung2006}. 
 \begin{figure}[!h]
 \centering
  \includegraphics[viewport = 120 70  580 550, scale=0.55,clip]{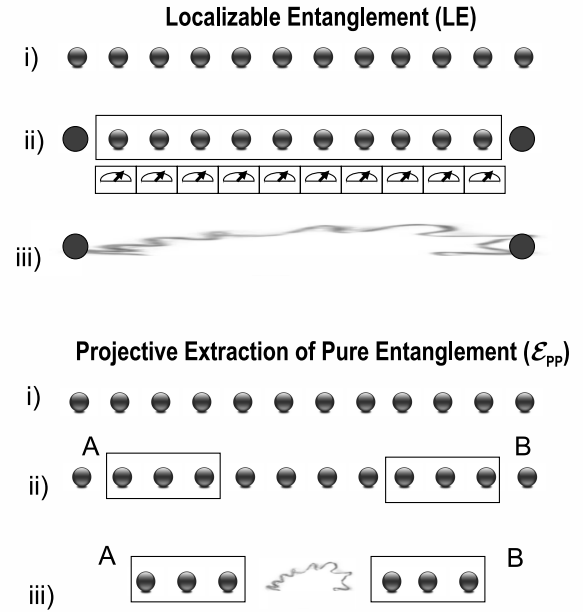}
 \caption{\textit{Localizable Entanglement} (Top). i) Representation of a quantum many body system. ii) To localize entanglement between the two end parties, we perform local measurements on the other parties. iii) LE is the maximum amount of entanglement that on average can be localized on these end parties by this procedure. \textit{Projective Extraction of Pure Entanglement} (Bottom). i) Representation of a quantum many body system. ii) We want to establish a pure entangled state between regions A and B by means of two coarse grained projective measurements on these regions. iii) The procedure can establish pure entangled states between regions A and B with some probability of success.\label{Fig1.}}
 \end{figure} 

In considering the use of many body systems for quantum information applications, one would ideally like to establish pure entangled states between distant regions, because all quantum protocols such as teleportation \cite{Bennett1993}, work best for pure states. There is also the additional technical advantage that one can use the von Neumann entropy of entanglement \cite{Popescu1992}, which is a very good measure for pure states to compute the entanglement. Indeed, for ground states which are pure, von Neumann entropy has been extensively used to calculate entanglement between complementary parts \cite{Xu2008}, \cite{Its2005}, \cite{Vidal2003}. In this context the amount of entanglement obtainable in useful form from these complementary parts for a single copy of a many-body state has also been discussed \cite{SingleCopy}. But for generic distant parts of a QMB system the state is a mixture, so a conversion to a pure state would be ideal. It would be desirable to achieve this by only local operations on those parts. Of course, in principle, one could use several copies and employ purification protocols, which require local actions as measurements and unitary operations as well as classical communication between the parties, to achieve a smaller number of almost pure states. However, it is highly desirable that these local operations be simple (such as projections) acting on a single copy of an entangled state.

In \cite{Popp2005} was introduced a new measure of bipartite entanglement called \textit{Localizable Entanglement} (LE) that
is operationally defined as the maximum amount of entanglement that can be localized, on average, between two spins of the chain by performing local measurements on other parties. Studies on the amount of LE as a function of the distance of the spins to be entangled and the influence of anisotropic couplings were conducted in \cite{Jin2004}.By definition, the LE procedure requires a fine grained access to all the subsystems of the many body system. This requirement seems rather difficult for typical situations where the subsystems are very close to each other. 

In this paper we present a procedure that, in a probabilistic sense, allows us to establish {\it pure} bipartite entangled states between distant regions of many body system by means of {\it local} projections {\it only} on these regions (Fig1). 

\section{Projective Extraction of Pure Entangled States }\label{sec.E_PP}

 We consider the state $|\phi\rangle$ of a system $\mathcal{S}$ defined on a lattice and the associated density operator $\rho_{0}=|\phi\rangle \langle \phi|$. We can define projective operators  $P_{A}$, $P_{B}$ acting locally on separate regions A and B of $\mathcal{S}$, where each of them may comprise several lattice sites. The state after local projection $P_{A} P_{B}$ is

\begin{equation}\label{eq.post-measurement}
\rho=\frac{(P_{A} P_{B}) \rho_{0} (P_{A}  P_{B})}{Tr(P_{A} P_{B}\rho_{0})}
\end{equation}

By an appropriate choice of the projectors $P_{A}$, $P_{B}$ the state $\rho_{AB}=Tr_{(\mathcal{S}-(A \cup B))}(\rho)$ will be pure, and the probability of obtaining this state when performing the associated selective measurement is $Prob(\rho_{AB})=Tr(P_{A} P_{B}\rho_{0})$. Once this pure state is established, the entanglement between regions A and B can be quantified by the von Neumann entropy $S(\rho_{A})=-Tr(\rho_A\log_2\rho_A)$ with $\rho_{A}=Tr_{B}(\rho_{AB})$. It is convenient to introduce
\begin{equation}
\mathcal{E}_{PP}(P_A,P_B)=S(\rho_{A}) \,Prob(\rho_{AB})
\end{equation}
where the subscript letters $PP$ indicate the features projective and pure. This quantity both captures the probabilistic aspects of this procedure and quantifies the amount of quantum correlations in the resulting state.
  The procedure is meaningful only for a choice of the projectors $P_A$ and $P_B$ that are  (i) suitable for  extraction of pure states and (ii) of rank higher than one because otherwise one would always obtain an unentangled state.
We shall make these requirements more concise in Section \ref{sec.genproc}.

\section{Some simple examples of $\mathcal{E}_{PP}$}
In this section we illustrate how the projective procedure can be applied in two simple examples.
\subsection{Supersinglets}\label{super}
Supersinglets are defined as \cite{Cabello2003}
\begin{equation}\label{eq.supers}
\ket{S^{(d)}_{N}}=\frac{1}{\sqrt{N!}}\sum\limits_{s_{1\ldots N}} \epsilon_{s_1,\ldots,s_N}\ket{s_1,s_2,\ldots,s_N}~,
\end{equation}
describing a highly entangled state of a number $N$ qudits with individual number $d$ of degrees of freedom. The summation runs over all permutations $s_{1\ldots N}$ of $N$-tuples of $s_l$, each of which assuming integer values $[1,d]$ and labeling the local basis states $\lbrace\ket{s_l}\rbrace$ on site $l$, and $\epsilon_{s_1,\ldots,s_N}$ denotes the generalized Levi-Civita tensor. It was recently found that supersinglet states also arise as ground states for permutation Hamiltonians of systems where $d=N$ (generalization of two spin-$\frac12$ coupled by a Heisenberg interaction) \cite{Hadley2008}.
These states are particularly well suited to illustrate the concept of pure entangled state extraction by local projections, as introduced in the previous Section. To this end we choose a simple system of three qutrits ($N=d=3$),(labelled A, B and C) the supersinglet state reads
\begin{equation}
 \ket{\Psi}=\frac{1}{\sqrt{3}}(\ket{1}_C\ket{\Phi_{2,3}}_{AB}+\ket{2}_C\ket{\Phi_{3,1}}_{AB}+\ket{3}_C\ket{\Phi_{1,2}}_{AB})
\end{equation}
where we defined $\ket{\Phi_{i,j}}\equiv\frac{1}{\sqrt{2}}(\ket{i}\ket{j}-\ket{j}\ket{i})$.
By choosing, e.g.
\begin{equation}
P_{A/B}=\ket{2}\bra{2}+\ket{3}\bra{3}
\end{equation}
after projection the qutrits A and B will be left in the pure state $\ket{\Phi_{23}}$ 
for which $S(\Phi_{23})=1$. In a corresponding selective measurement, this will happen with probability $\frac13$. Then we find $\mathcal{E}_{PP}=\frac13$. To hightlight requirement (ii), we stress that by choosing rank-1 projector, e.g.
\begin{eqnarray}\nonumber
P_{A}=\ket{2}\bra{2}\\
P_{B}=\ket{3}\bra{3}
\end{eqnarray}
the state after projection will be pure but separable, namely $\ket{2}_A\ket{3}_B$. Rank-3 projectors for a qutrit act trivially as the identity, and leave the subsystems A and B in a maximally mixed state. For general $d=N$ one finds $\mathcal{E}_{PP}=\frac{1}{N}$~.

\subsection{$\mathcal{E}_{PP}$ in a non-equilibrium scenario}\label{locquench}
In this Section, we consider a ring of $N$ spins interacting through the exchange Hamiltonian
\begin{equation}\label{hamiltonian1}
\mathcal{H}_{0}=-\sum_{i=1}^{N}(\sigma^{x}_{i}\sigma^{x}_{i+1}+\sigma^{y}_{i}\sigma^{y}_{i+1})
\end{equation}
with periodic boundary conditions $\sigma_{N+1}^{\alpha} = \sigma_{1}^{\alpha}$ $\alpha = x,y$, where $\sigma^{x}$ and $\sigma^{y}$ are the Pauli $\sigma$ matrices. The initial state of the system is not an eigenstate of (\ref{hamiltonian1}) and consists of two spin flips located at sites $r_{1}$ and $r_{2}$ of the  ring
\begin{equation}\label{initState}
|\phi\rangle=\frac{1}{\sqrt{2}}(|r_{1}\rangle|r_{2}\rangle - |r_{2}\rangle|r_{1}\rangle)
\end{equation}

This physical situation allows temporal evolution for (\ref{initState}) giving the state for certain time $t$, $\ket{\phi_{t}}=e^{-i\mathcal{H}_{0}t}\ket{\phi}$, 
\begin{equation}\label{evolvedState}
\ket{\phi_{t}}=\frac{1}{\sqrt{2}}\sum_{s_{1},s_{2}=1}^N \epsilon(s_{2}-s_{1})A_{s_{1},s_{2}}(t)\ket{s_{1}} \ket{s_{2}}
\end{equation}
where $A_{s_{1},s_{2}}(t)=(f^N_{s_{1}r_{1}}(t)f^N_{s_{2}r_{2}}(t)-f^N_{s_{2}r_{1}}(t)f^N_{s_{1}r_{2}}(t))$ is the amplitude that at time $t$, one of the flips being at site $s_{1}$ and the other one being at site $s_{2}$ and 
\begin{equation}\label{amplitude}
f^N_{sr}(t)=e^{-i4t}i^d\sum_{k=-\infty}^{+\infty}J_{d-kN}(\beta)i^{-ikN}e^{(i2\pi fk)}
\end{equation}

is the amplitude for one of the flips to be initially at $r$ and at time $t$ be located at $s$, with $\beta = 4t$, $d=s-r$, $f=0.5$ and $\epsilon(i-j)$ is the sign function ($\epsilon(x)=+1, x>0; \epsilon(x)=-1, x<0; \epsilon(x)=0, x=0)$ and $J_{\nu}(x)$ is a Bessel function of the first kind.

 \begin{figure}
 \includegraphics[scale=0.325]{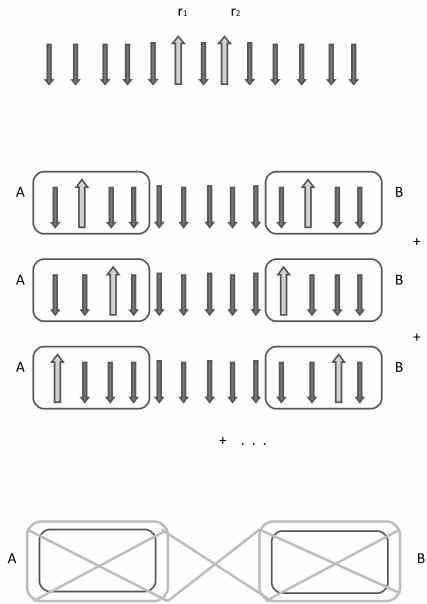}
 \caption{Schematic diagram of the $\mathcal{E}_{PP}$ in the XY ring with two spin flips. The initial state of consists in two localized flips in an initially non-entangled state (Top). After some time evolution, there is certain probability that one of the spin flips lies within the region A and the other one lies within the region B. By projecting the evolved state by means of two coarse grained measurements on these regions (Middle), an entangled pure state between A and B can be established when the results of the measurements ascertain that one spin flip is in A and the other one in B while saying nothing about their exact localization within A and B (Bottom). \label{Fig2.}}
 \end{figure}

The objective of our procedure is to create a pure entangled state between two separate regions A and B of the ring by projecting (\ref{evolvedState}) with two local coarse grained  measurements which ascertain whether there is some flip present in certain region without revealing any information about its position within the region (see illustration in Fig.~\ref{Fig2.}). This procedure discards trials in which no flips are found in any region or only one of the flips is found in one the measurement regions. The projected state after measurement is the result of selecting the cases in which we find one flip in A and the other one in region B. This setting of our projection procedure  can be made more concise by explicitely writting the projectors as  $P_{A}=\sum_{j \in A}\ket{j}\bra{j}$ and $P_{B}=\sum_{k \in B}\ket{k}\bra{k}$. So, taking into account (\ref{evolvedState}) and (\ref{amplitude}) the state after projection $\ket{\Psi_{t}}_{AB}=P_{A}P_{B}\ket{\phi_{t}}$ can be written as

\begin{equation}\label{projState}
\ket{\Psi_{t}}_{AB}=\frac{1}{\sqrt{2}}\sum_{j \in A, k \in B}\epsilon(k-j)A_{jk}(t)\ket{j}\ket{k}
\end{equation}
with  $A_{jk}(t)=(f_{j1}(t)^{N}f_{k2}(t)^{N}-f_{j2}(t)^{N}f_{k1}(t)^{N})$ and also noting that $\forall j \in A, k \in B \Rightarrow \epsilon(k-j)=+1$.

\begin{figure}
 \includegraphics[scale=0.55]{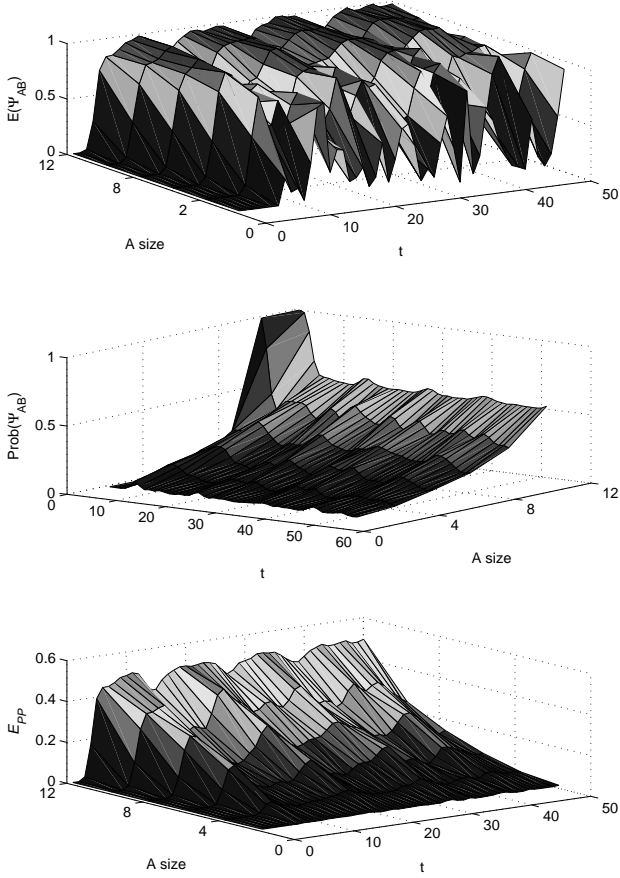}
 \caption{Entanglement Entropy {\bf(top)}, probability of success, {\bf(middle)}, and $\mathcal{E}_{PP}$ {\bf(bottom)} of the system of two spin flips in the XY ring  as function of time and size of the measurement areas A and B on the ring. Entanglement, probability of success and $\mathcal{E}_{PP}$ scale as the size of areas increase. When the size of the measurement areas equal the half of the chain, $\mathcal{E}_{PP}$  slightly oscillates around the value of 0.5. \label{Fig3.}}
 \end{figure}

In order to compute the $\mathcal{E}_{PP}$ resulting for our particular choice of projectors $P_{A}$ and $P_{B}$ we have to compute the von Neumann entropy of the reduced density matrix of one of the regions, that in the basis of lattice sites (real space basis) it is easy to show that has the form, $(\rho_{A}(t))_{j,j^{'}}=\bra{ j^{'}}\rho_{A}(t)\ket{j} = \sum_{k \in B} A_{jk}(t)A^{*}_{j^{'}k}(t)$ where $j$ and $j^{'}$ $\in A$. The probability for obtaining the state (\ref{projState}) in our procedure also easily reads as
\begin{equation}\label{probability}
 Prob(\ket{\Psi_t}_{AB})= \frac{2\sum_{j \in A, k \in B} |A_{jk}(t)|^{2}}{\sum_{l,m \in \textit{R}}  |A_{lm}(t)|^{2}}
\end{equation}
where \textit{R} refers to all sites in the ring. This probability can be interpreted as the probability of success of the procedure.

Fig 3. shows the results of the simulations carried out in this \textit{out of equilibrium} setting. We have used a ring with $N=24$ spins and the spin flips are initially located at $r_{1}=10$ and $r_{2}=14$. As the size of measurement regions increase in the number of sites they contain, the probability of success and the $\mathcal{E}_{PP}$ also increase. In the limit in which regions A and B equal the half of the chain, the probability of success $ Prob(\ket{\Psi_t}_{AB})=0.5$  and the value of $\mathcal{E}_{PP}$ slightly oscillates around $1/2$. These values are decreasingly modulated as the size of  areas are concurrently decreased. The time evolution induces some oscillatory effects in the values for entanglement, probability and $\mathcal{E}_{PP}$ for a fixed size of the measurement areas. For the type of system studied in this section and the type of projections we made, $\mathcal{E}_{PP}$ turns out to be equivalent to the notion of entanglement of particles of Refs \cite{Wiseman2003}, \cite{Dowling2006}. In \cite{Molina2008} this procedure has been applied to the continuous case of 1D impenetrable bosons lying on a ring and authors have also shown the scaling of the $\mathcal{E}_{PP}$ with the number of particles in the ring. An interesting open question is the issue of measuring the entanglement created by our procedure.

\subsection{$\mathcal{E}_{PP}$ in the ground state of a QMBS} \label{sec.groundstate}
We choose to illustrate this section, the ground state of the ring of spins in an external field $h$ interacting through the Hamiltonian 
\begin{equation}\label{hamiltonian2}
\mathcal{H}=\mathcal{H}_0- 2 h S^z 
\end{equation}
where $S^z=\sum_{i=1}^N \sigma_i^z$. Noting that $[\mathcal{H}_0, S^z]=0$, the eigenvectors of $\mathcal{H}$ are determined by a positive integer $m\leq N$ and a collection of real numbers $\{p_i\}_{i=1}^m$, with each $-\pi<p_i<\pi$, called momenta \cite{Korepin1993}.  For short, the collection $\{p_i\}_{i=1}^m$ may be denoted simply by $\{p\}$.  Thus eigenvectors of (\ref{hamiltonian2}) are defined by the set of vectors $\ket{\Phi_m\{p\}}$ with


\begin{equation}\label{eigenvectors}
\ket{\Phi_m\{p\}}=\frac{1}{\sqrt{m!}}\sum_{x_1, \ldots, x_m = 1}^{N}\chi_m(\{x\}|\{p\})\ket{x_{m} \cdots x_{1}}
\end{equation}
The complex valued function $\chi$ is defined by
\begin{equation}\label{wavefunctions}
\chi_m\left(\{x\}|\{p\}\right)= \left( \prod_{1 \leq a < b \leq m} \epsilon(x_b-x_a)\right) \det (D_{\{x\}|\{p\}})
\end{equation}
where $(D_{\{x\}|\{p\}}$ is the $m\times m$ matrix with $(j,k)$ entry $D_{jk}=\exp(ix_j p_k)$. The function $\chi(\{x\}|\{p\})$ is called a wave function and it is symmetric in $x$ and antisymmetric in $p$.

 
For some high values of the magnetic field $h$ in (\ref{hamiltonian2}) the ground state of the ring can be expressed as the $m=2$ case of the eigenvector (\ref{eigenvectors}), i.e.,
\begin{equation}\label{eq.twoflip}
 \ket{\Phi_{2}\{p\}}=\frac{1}{\sqrt{2}}\sum_{\{x\}=1}^N \epsilon(x_{2}-x_{1}) det(D_{\{x\}|\{p\}}) \ket{\{x\}} 
\end{equation}
where $\{x\}= (x_1,x_2)$, $\ket{\{x\}} = \ket{x_1 x_2}$ and $det(D_{\{x\}|\{p\}})$ means the determinant of the  $D_{2 \times 2}$ matrix with elements $D_{in}=e^{(ip_{n}x_{i})}$ with $\{p\}$ being the momenta of the two spin flips. 
We choose the same projectors $P_{A}$ and $P_{B}$ for the \textit{Pure state Entanglement Extraction} as the ones proposed in the previous Section. The bipartite projected pure state $\ket{\Psi}_{AB} = P_{A}P_{B}\ket{\Phi_{2}\{p\}}$ can be read now as,
\begin{equation}\label{projState2}
 \ket{\Psi}_{AB}=\frac{1}{\sqrt{2}}\sum_{x_{j} \in A,x_{k} \in B} \epsilon(x_{k}-x_{j}) det(D_{\{x_{j},x_{k}\}|\{p\}}) \ket{x_{j}} \ket{x_{k}}
\end{equation}  
noting that now $\epsilon(x_{k}-x_{j})=+1$ $ \forall j \in A, k \in B$. As in the previous section the reduced density matrix of region A can be described by $\bra{ j^{'}}\rho_{A}\ket{j} = \frac{1}{2} \sum_{k \in B} A_{jk}A_{j^{'}k}^{*}$ where $A_{jk}=\epsilon(x_{k}-x_{j})det(D_{\{x_{j}, x_{k}\}|\{p\}})$. Fig 4 shows the values of $\mathcal{E}_{PP}$, $Prob(\ket{\Psi}_{AB})$ and the entanglement of the projected state $E(\ket{\Psi_{AB}})$ as the size of the measurement region A (B) increases.

\begin{figure}[ht]
\centering
 \includegraphics[scale=0.5]{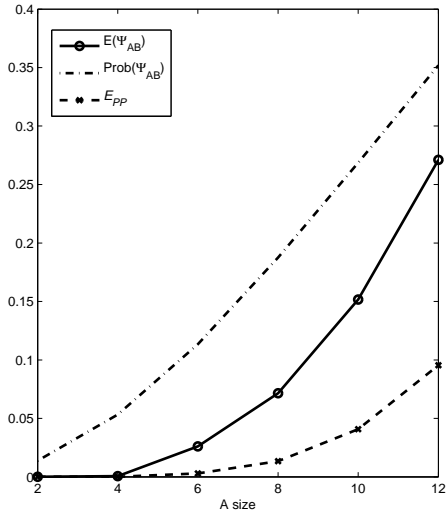}
\caption{Entanglement entropy, probability of success and $\mathcal{E}_{PP}$ in the  ground state of the ring ($N=24$) with Hamiltonian (\ref{hamiltonian2}) as the number of sites comprising the measurement region A varies from $\Delta_{A}=2$ $\to$ $\Delta_{A}=N/2$. \label{Fig4.}}
\end{figure}

\section{General Procedure for Extracting Pure Entangled States by Distant Local Projective Measurements}
In contrast to the preceeding Sections, here we elaborate on whether one can find nonzero $\mathcal{E}_{PP}$ in situations where the properties of ground state of a QMBS (symmetries, good quantum numbers) do not a priori suggest a heuristic choice for the suitable projectors for the projective extraction procedure. To this end we provide a general prescription that allows to systematically quantify the maximally achievable $\mathcal{E}_{PP}$ for a given state, when the distant parties A and B have access to a fixed number of contiguous sites.

We begin with specifying, under which circumstances a local projective measurement on subsystems of a MBS can lead to a pure state. I.e. we ask for the criterion the projector $P=P_A P_B$ must satisfy, so that
\begin{equation}\label{eq.purify}
 P \rho_{AB} P \rightarrow \ket{\Phi}\bra{\Phi}
\end{equation}
For this purpose it is instructive to consider the reduced density matrix of the two separate regions A and B $\rho_{AB}$ in its diagonal representation (Schmidt representation, assuming the ground state is pure)
\begin{equation}\label{reddens}
\rho_{AB}=\sum_{j}\lambda_{j}\ket{\Psi_{AB}^{j}}\bra{\Psi_{AB}^{j}}
\end{equation}

There could be two alternative ways of fulfilling Eq.~(\ref{eq.purify}), being a) by finding a $P=P_j$ so that
\begin{eqnarray}\label{poss_A1}
P_j\ket{\Psi^k}&=&0\quad j\neq k \\\label{poss_A2}
P_j\ket{\Psi^j}&=&\ket{\Phi^j}
\end{eqnarray}
resulting in pure state $\ket{\Phi}=\ket{\Phi^j}$ after projection or b) by having a set of at least two Schmidt vectors $\lbrace\ket{\Psi^\alpha}\rbrace$ which after projection by $P$ result in
\begin{eqnarray}\label{poss_B}
P\ket{\Psi^\alpha}&=&\ket{\Phi}\; \forall\; \alpha.
\end{eqnarray}
It is easy to show that one can exclude possibility b) by virtue of the orthogonality $\langle\Psi^k\ket{\Psi^l}=\delta_{k,l}$ and the idempotence of the projectors $P^2=P$.

In order to construct a trial set of local projectors $P_A$ and $P_B$, we proceed by casting the vectors $\ket{\Psi^j}$ into the form
\begin{equation}\label{2ndSchmidt}
\ket{\Psi^{j}}=\sum_{i}\omega_{i}^{j} \ket{\alpha_{i}^{j}} \ket{\beta_{i}^{j}} 
\end{equation}
which is the Schmidt decomposition of $\ket{\Psi^j}$.
We require the global projectors $P_j$ to be of direct product type $P_j=P^j_A\otimes P^j_B$ and fulfill the condition a) (Eqs.~(\ref{poss_A1}),(\ref{poss_A2})). Representation Eq.~(\ref{2ndSchmidt}) suggests the following construction prescription
\begin{align}\label{Projectors_Alg}
P^{j\mu}_{A}&=\sum_{k \in \mathrm{I}_\mu} \ket{\alpha_{k}^{j}}\bra{\alpha_{k}^{j}}  \\ \nonumber
P^{j\nu}_{B}&=\sum_{m \in \mathrm{I}_\nu}\ket{\beta_{m}^{j}}\bra{\beta_{m}^{j}} .
\end{align}
The additional superscripts $\mu,\nu$ label different sets of indices $\mathrm{I}_{\mu,\nu}$ encoding the the $\mu,\nu$-th combination of rank-1 projectors (e.g., $\ket{\alpha_{k}^{j}}\bra{\alpha_{k}^{j}}$), leading to rank $\geq 2$ (see also Sec.~\ref{super}). E.g. for some particular $\nu=3$, $\mu=4$ and $\mathrm{I}_3=\lbrace 3, 5 \rbrace$, $\mathrm{I}_4=\lbrace 2, 7 \rbrace$ the projectors in Eq.~\ref{Projectors_Alg} would become
\begin{align}\label{Projectors_Alg_example}\nonumber
P^{j3}_{A}&= \ket{\alpha_{3}^{j}}\bra{\alpha_{3}^{j}} + \ket{\alpha_{5}^{j}}\bra{\alpha_{5}^{j}}  \\ \nonumber
P^{j4}_{B}&= \ket{\beta_{2}^{j}}\bra{\beta_{2}^{j}} + \ket{\beta_{7}^{j}}\bra{\beta_{7}^{j}}~.
\end{align}
We then proceed by systematically checking for the existence of projectors $P^{j (\mu \nu)}=P^{j\mu}_A\otimes P^{j\nu}_B$ being suitable for the extraction of pure states. If several such projectors can be found, the corresponding extracted states $\ket{\Phi}^{j (\mu \nu)}$ will be evaluated for their entanglement content as measured by the von Neumann entropy $S( Tr_B( \ket{\Phi}^{j (\mu \nu)} \bra{\Phi} ) )$. $\mathcal{E}_{PP}$ is then defined by the maximally achievable amount $Prob(\ket{\Phi}^{j (\mu \nu)}))\,S(Tr_B(\ket{\Phi}^{j (\mu \nu)}\bra{\Phi}))$ among these states. In the present notation the probabilities of obtaining the states read 
\begin{equation}
 Prob(\ket{\Phi}^{j (\mu \nu)}))=\lambda_j\,\bra{\Psi^{j}}P^{j (\mu \nu)}\ket{\Psi^{j}}
\end{equation}

As for our procedure Eqs.~(\ref{poss_A1}) and (\ref{poss_A2}) express the only possible scenario in which pure states can be extracted, $\mathcal{E}_{PP}$ will not merely be a lower bound but quantify the optimal performance of extracting pure state entanglement by local projections.
\subsection{Application of the general procedure: $\mathcal{E}_{PP}$ for a general phase of a spin model}\label{sec.genproc}
 \begin{figure}
 \centering
  \includegraphics[scale=0.75]{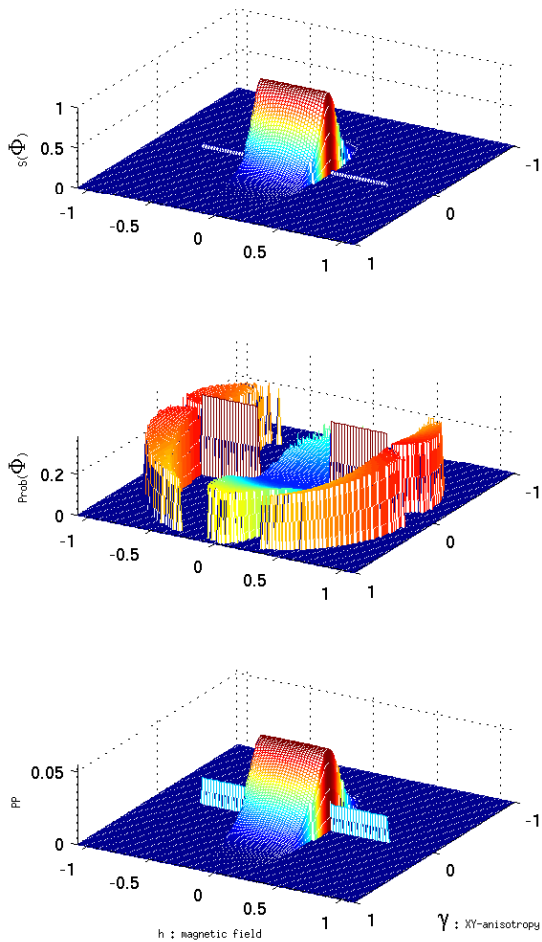}
 \caption{\textit{General procedure} {\bf (top)} Maximum entanglement entropy {\bf (middle)} Maximum probability of extracting a pure state {\bf (bottom)} projectively extractable pure entanglement}\label{fig.genproc}
 \end{figure} 

For the study of more general instances of $\mathcal{E}_{PP}$ we now slightly alter the type of Hamiltonian used so far, in that we introduce an anisotropy causing the model to have less symmetries and therefore less good quantum numbers.
The transverse XY Hamiltonian reads
\begin{equation}\label{xyhamiltonian}
\mathcal{H}_{XY}=-\sum_{i=1}^{N}\left((1-\gamma)\sigma^{x}_{i}\sigma^{x}_{i+1}+(1+\gamma)\sigma^{y}_{i}\sigma^{y}_{i+1}\right) - 2 h S_z
\end{equation}
We shall explore the different ground states of $\mathcal{H}_{XY}$ for a particular system of $N=6$ spins arranged on a ring, partitioned into symmetric regions A and B of two contiguous spins each that are separated by a single site.
For $\gamma\neq 0$ the z-Magnetization is no longer conserved, leaving only the parity $\mathit{P}_z=\prod_{k=1}^N\, \sigma^z_k$ as a commuting observable. We shall now apply our general procedure and evaluate $\mathcal{E}_{PP}$ as a function of parameters $h$ and $\gamma$, transverse field and anisotropy respectively.
Figure \ref{fig.genproc} shows that nonzero pure entanglement can indeed be extracted by local projections, even outside the previously covered regimes, where the ground state is a superposition state of the form Eq.~(\ref{eq.twoflip}). In the present study, this corresponds to $\gamma=0$ and $h\simeq\pm 0.5 $~. Beyond these regimes our procedure reveals that pure entanglement can be extracted in a sector of the ($\gamma$,$h$) plane for which the ground states attain odd parity, i.e. $\langle\mathit{P}_z\rangle=-1$ , providing strong evidence that the origin of nonzero $\mathcal{E}_{PP}$, as detected by our general procedure, can be achieved by a local parity measurement. Let us examine under which circumstances locally measuring the parity in regions A and B can lead to a pure entangled state (for simplicity restricting ourselves to the isotropic $(\gamma=0,h=0)$ case, where the number of spin flips $m=\frac{N}{2}$ is conserved). An important role must be attributed to the particular partitioning we chose above, i.e. the complement of the measurement regions is comprised of merely two sites. If the parity of regions A and B is even $(\ket{\downarrow\downarrow},\ket{\uparrow\uparrow})$, then the complementary spins must be of odd parity $(\ket{\uparrow\downarrow},\ket{\downarrow\uparrow})$, to comply with the global odd parity of ground state. The local measurements of parity must not discriminate between the particular configurations which is a necessary requirement of having an entangled state after projection, and is equivalent to the aforementioned necessity of having local projectors of rank higher than one. Then, measuring even parity in both regions A and B ($P_A=P_B=\ket{\uparrow,\uparrow}\bra{\uparrow,\uparrow}+\ket{\downarrow,\downarrow}\bra{\downarrow,\downarrow}$) is consistent with four possible spin configurations, which occur with equal amplitudes in the ground state, thus after projection the state of the whole system reads $\ket{\psi}=\frac{1}{\sqrt{2}}(\ket{\uparrow\downarrow}+\ket{\downarrow\uparrow})_{\overline{AB}}\otimes(\ket{\uparrow\uparrow}_A\ket{\downarrow\downarrow}_B+\ket{\downarrow\downarrow}_A\ket{ \uparrow\uparrow}_B)$ which was verified numerically. The state of regions A,B is therefore pure and entangled, which is clear by inspection. This study therefore sheds light on the interrelation between $\mathcal{E}_{PP}$ and observables that commute with the underlying Hamiltonian, and further have a degenerate spectrum. We stress that our study has primarily been conducted, at the current stage, for conceptual purposes. Whether nonzero $\mathcal{E}_{PP}$ exists in more realistic or potentially useful settings can not be said at this point.

\section{Discussion}
The non equilibrium setting of Section \ref{locquench} shows qualitative similarities to the features after a \textit{local quench} (\cite{White2004}, \cite{Calabrese2007}, \cite{Eisler2008}). 
Calabrese and Cardy recently proposed \cite{Calabrese2005} a simple interpretation for the observed behavior in terms of quasiparticle excitations emitted from the initial state at $t=0$ that freely and semiclassicaly propagate with velocity $v$.
If the initial state is separable (in the standard basis) particles that originate from different sites are incoherent, but pairs of particles moving to the left or right from a given point are highly entangled. The reasoning of these authors is that a point $x_{A}$ in region $A$ will be entangled with a point $x_{B}$ in region $B$ if a pair of entangled particles emitted at an earlier time arrive simultaneously at $x_{A}$ and $x_{B}$. This picture of a {\em causal cone} is capable of qualitatively explaining the observed saturation of the block entropy of contiguous blocks of spins in an infinite chain after the quench \cite{Calabrese2007}.
It is thus tempting to qualitativeley interpret the results of Section (\ref{locquench}) where a saturation is observed (for a time window before finite size effects kick in).
Our procedure of selecting particular measurements outcomes assures that these excitations have arrived to regions $A$ and $B$ at the instant of measurement which, in agreement with the quasiparticle picture, may lead to entanglement after a time that scales linearly with the distance between these regions due to the finite speed for the propagation of the excitations $v$. More importantly such a measurement does not discriminate particular points $x_{A}$ and $x_{B}$ within these regions, and therefore the quantum indeterminacy (owing to the time evolution) is what gives us the possibility for establishing a pure entangled state between $A$ and $B$. The growth and later saturation of entanglement entropy between blocks $A$ and $B$ can be viewed from the fact that due to time evolution, there is an instant $t^{*}$ at which there is a nonzero and non increasing probability that the excitations have arrived to all the points $x_{A}$ and $x_{B}$ lying within regions $A$ and $B$.

As particularly highlighted in Sec \ref{sec.genproc} a vital aspect of $\mathcal{E}_{PP}$ is its connection to good quantum numbers, which may lead to a similar quantum indeterminacy as mentioned for the dynamic case above. This can arise for the ground state of a QMBS if the underlying Hamiltonian commutes with an observable that has a degenerate spectrum, like in our cases the magnetization or parity. If these observables can also be measured locally, and the local measurement outcomes are again consistent with more than one global state configuration (degeneracy), we have shown that this indeterminacy can be exploited in that it may lead to pure entangled states of different regions of the MBS after the measurement. So we are deliberately measuring some {\em incomplete} set of commuting observables locally, which will maintain the (quantum) uncertainty of telling in which particular state the subsystem resides is maintained. As these local measurements are performed on both separated regions, the projected state has to accomplish both local degeneracies simultaneously which is the origin of the nonlocal quantum correlation.
Under which circumstances this quantum correlation is that of a pure state is connected rather subtly to an interplay of the observable being measured, the partitioning of the system into measurement regions A,B and their complement and the global properties of the state.

Another interesting point to discuss is the topic related with the transferring of entanglement created by our procedure to systems in which this entanglement could be used as a resource for quantum information tasks. In other words, what we would like to do is to be able to swap the entanglement established between regions $A$ and $B$ onto a couple of related systems $\mathcal{S}_{A}$ and $\mathcal{S}_{B}$. In \cite{Retzker2005}, authors propose a protocol by which the entanglement between separated regions in an ion trap could be transferred to a pair of ancillary ions each one of them related with each one of the regions.
Further investigations should clarify if the above kind of operations feasibly allow to use the entanglement extracted by our procedure in quantum information tasks.

\begin{acknowledgments}
JMV acknowledges the supporting of Spanish Office for Science and
Technology program ``Jose Castillejo'' and Fundacion Seneca Murcia. The studentship of HW is supported by the EPSRC, UK. SB acknowledges the EPSRC, UK, the EPSRC sponsored QIPIRC, the Royal Society and the Wolfson Foundation.
\end{acknowledgments}


\end{document}